%% file: main.tex
\documentclass[Times2COL]{WileyNJDv5}
\usepackage{config}

\articletype{Article Type}%

\received{00}
\revised{00}
\accepted{00}
\journal{Journal}
\volume{00}
\copyyear{2023}
\startpage{1}

\begin{document}

\title{ECLYPSE: a Python Framework for Simulation and Emulation of the Cloud-Edge Continuum}

\author[1,2]{Jacopo Massa}
\author[1]{Valerio De Caro}
\author[1]{Stefano Forti}
\author[1]{Patrizio Dazzi}
\author[1]{Davide Bacciu}
\author[1]{Antonio Brogi}

\authormark{Massa \textsc{et al.}}
\titlemark{ECLYPSE: a Python Framework for Simulation and Emulation of the Cloud-Edge Continuum}

\address[1]{\orgdiv{Department of Computer Science}, \orgname{University of Pisa}, \orgaddress{\city{Pisa}, \country{Italy}}}
\address[2]{\orgdiv{ISTI-CNR}, \orgaddress{\city{Pisa}, \country{Italy}}}

\corres{Corresponding author: Jacopo Massa \email{jacopo.massa@di.unipi.it}}



\abstract[Abstract]{The Cloud-Edge continuum enhances application performance by bringing computation closer to data sources. However, it presents considerable challenges in managing resources and determining application service placement, as these tasks require analysing diverse, dynamic environments characterised by fluctuating network conditions. Addressing these challenges calls for tools combining simulation and emulation of Cloud-Edge systems to rigorously assess novel application and resource management strategies.
In this paper, we introduce \eclypse, a Python-based framework that enables the simulation and emulation of the Cloud-Edge continuum via adaptable resource allocation and service placement models. \eclypse features an event-driven architecture for dynamically adapting network configurations and resources. 
It also supports seamless transitions between simulated and emulated setups, thus enabling the execution of experiments in simulated, emulated, and hybrid settings. In this work, we illustrate and assess \eclypse capabilities over three use cases, showing how the framework supports rapid prototyping across diverse scenarios.
}



\maketitle

\renewcommand\thefootnote{}

\renewcommand\thefootnote{\arabic{footnote}}
\setcounter{footnote}{1}

\input{src/introduction}
\input{src/model/main}
\input{src/framework/main}
\input{src/experiment/main}
\input{src/related}
\input{src/conclusions}
\input{src/acknowledgments}






\bmsection*{References}\vspace{2mm}
\AtNextBibliography{\footnotesize}
\printbibliography[heading=none]



\end{document}

%% file: src/introduction.tex
\section{Introduction}
\label{sec:introduction}

\noindent
The rapid evolution of Cloud-Edge paradigms has significantly broadened the landscape of distributed computing, enabling a wide range of applications in sectors such as real-time analytics, urban infrastructure, and autonomous systems~\cite{rosendo202271}. By enabling computation to occur closer to data sources, the Cloud-Edge continuum enhances responsiveness, scalability, and energy efficiency across heterogeneous and dynamic environments. However, this paradigm introduces considerable complexity in terms of resource management and application service placement, as applications must adapt to fluctuating network conditions, variable resource availability, and evolving user demands~\cite{kimovski2022mobility, zeydan2021intelligent}.

Optimising placement and workload allocation in such settings requires sophisticated decision-making processes that account for multiple constraints -- including latency, bandwidth, hardware limitations, and user location. Static or monolithic models are insufficient to capture the dynamic and context-sensitive nature of real-world scenarios~\cite{luckow2021exploring}. Consequently, flexible frameworks are needed to support adaptive, reproducible experimentation at scale, possibly without leasing expensive computing resources from various providers.

To address this, simulation offers a practical and controlled environment to evaluate theoretical models under large-scale configurations. Yet, while simulations offer predictive value, they often lack the fidelity to capture the nuanced behaviour of real systems. Emulation bridges this gap by enabling applications to execute under near-realistic conditions, thereby providing insights into operational behaviour that simulations alone cannot~\cite{zeng2019emuedge}. Hybrid approaches that integrate simulation with emulation have thus emerged as a promosing strategy to explore performance, robustness, and adaptability of Cloud-Edge systems.

Despite the availability of various simulators and emulators~\cite{survey2019FI,survey2020_DeMaio,survey2020_Kertesz}, a number of limitations persist. Many tools lack flexibility, extensibility, or adequate documentation~\cite{usability}, limiting their utility for:
\begin{enumerate*}[label=\textit{(\roman{*})}]
    \item validating placement and resource allocation strategies in dynamic environments,
    \item testing innovative service designs, particularly those involving machine learning or adaptive workloads, and
    \item conducting cost-effective experiments without relying on physical infrastructure.
\end{enumerate*}

In response to these limitations, in this article, we present \eclypse\footnote{\url{https://github.com/eclypse-org/eclypse}}, a Python-based framework designed to seamlessly combine simulation and emulation for the development, testing, and optimisation of Cloud-Edge applications. \eclypse adopts an event-driven architecture built on a formal environment model, offering precise control over the simulation dynamics and allowing fine-grained orchestration of runtime behaviour.

A key advantage of \eclypse lies in its Python implementation, which brings several concrete advantages. Python's simple and expressive syntax facilitates rapid prototyping and ease of use, even for non-specialists -- an aspect confirmed by comparative analyses showing that Python excels in readability and writability among common programming languages~\cite{ahmed2021comparative}. In addition, Python's vast ecosystem of domain-specific libraries enables seamless integration with advanced workloads and machine learning frameworks~\cite{pytorch,tensorflow,scikit-learn}. Its cross-platform compatibility, combined with performance enhancements via tools like \cd{Cython}, allows for efficient execution even in resource-constrained scenarios. Finally, being available as a standard \cd{pip} package, \eclypse ensures accessibility, maintainability, and ease of adoption.
The objectives of this paper are threefold, namely:
\begin{enumerate}[label=\textit{(\roman{*})}]
    \item introducing a unified simulation/emulation framework for the Cloud-Edge continuum,
    \item demonstrating its flexibility and extensibility across diverse experimental scenarios, and
    \item illustrating how Python's ecosystem and modular design support realistic, reproducible, and cost-efficient experimentation.
\end{enumerate}

The rest of this article is structured as follows.~\cref{sec:model} presents the \eclypse environment model, describing assets, infrastructure, applications, and placement.~\cref{sec:framework} details the simulation and emulation architecture.~\cref{sec:experiments} reports experimental results from multiple use cases.~\cref{sec:related} reviews relevant literature and distinguishes \eclypse from existing tools. Finally,~\cref{sec:conclusions} concludes the paper and outlines directions for future work.

%% file: src/model/main.tex
\section{ECLYPSE Model}
\label{sec:model}


\input{src/model/environment}
\input{src/model/simulation}

%% file: src/model/environment.tex
\subsection{Environment Model}\label{sec:model_env}

The Cloud-Edge continuum is characterised by large scale and high heterogeneity, making it challenging to establish a single comprehensive model encompassing its complexity.
First, the set of modelled resources within an infrastructure may be highly diverse depending on the provider and the adopted cost models adopted. This also applies to applications, depending on the service they provide and the targets of their operators.
Finally, the purposes of simulating and emulating Cloud-Edge settings may be diverse. For instance, an infrastructure provider aims at optimising the use of resources, whereas an application developer may seek to validate an implementation on a specific infrastructure. Therefore, a flexible and precise model is essential to define the intended environment.

\input{fig/environment}

This section illustrates the formal environment model of \eclypse according to the top-down flow shown in~\cref{fig:environment}. 
\cd{Assets} allow quantifying and managing the resources needed to deploy services, ensuring that the right amount of each resource is available where and when it is required. 
Then, the \cd{Infrastructure} is the computation and communication substrate of the Cloud-Edge environment.
Finally, an \cd{Application} refers to a collection of services that interact in a structured manner, addressing infrastructure challenges by aligning asset definitions with corresponding requirements. The \cdp{Placement} represent this relationship by associating application services with the infrastructure nodes onto which they run.

\input{tab/notation}
\input{tab/assets}

\subsubsection{Asset}\label{def:asset}An asset $\asset$ denotes resource capabilities and requirements (\eg CPU, memory, bandwidth) and is defined as a triple $(\domain, \agg, \sat)$, where:
\begin{itemize}
    \item $\domain$ is the asset \textit{domain}, defining the set of all possible values the asset can take.
    \item $\sat: \domain \times \domain \to \{0, 1\}$ is a function that inherently establishes an order among values within the domain $\domain$. For any two elements $\ass_1, \ass_2 \in \domain$, the function evaluates them and outputs 1 if $\ass_1$ precedes $\ass_2$ according to the predefined ordering, and 0 otherwise. This ordering represents a \textit{comparison relation} between asset values, where $\lb$ denotes the lowest value in the domain (\ie $\lb \sat \ass\ = 1$ for every $\ass \in \domain$), and $\ub$ is the highest value in the domain (\ie $\ass \sat \ub = 1$ for every $\ass \in \domain$).
    \item $\agg: \domain \times \domain \to \domain$ denotes an \textit{aggregation} function, enabling the \textit{summing} of instances of identical types of assets to form a unified representative value.
\end{itemize}
Assets are classified according to their aggregation properties, and each type follows specific rules for comparison and aggregation.
We provide five categories covering a wide range of assets, summarised in \cref{tab:assets}: \textit{additive}, \textit{concave}, \textit{convex}, \textit{multiplicative}, and \textit{symbolic}. 
For instance, the number of CPUs can be represented as an additive asset. If an infrastructure node equipped with 64 CPU cores is upgraded by including an additional 64 cores, the total number of CPU cores is 128, resulting in better resource quality. Instead, latency is a concave asset, as a lower value characterises a better link. Finally, availability is a multiplicative asset, as its value lies in the interval $[0,1]$, and cumulative availability is obtained as the probability that all nodes are available simultaneously.

In the following, we ease the discussion by defining sets of multiple assets as \textit{buckets}, denoting them with $\assets$ and their (set of) values as $\bass$. The aggregation of values in a bucket is conducted element-wise, and the comparison yields 1 only if every element-wise comparison results in 1.
For instance, the RAM, CPU and storage of a node (or required by a service) constitute a bucket.

\subsubsection{Infrastructure}\label{def:infra}We define the infrastructure $\infra$ a 5-tuple $(\nodes, \links, \noderes, \linkres, \pathres)$, where:
\begin{itemize}
    \item $\nodes = \{(i, \bass_i)\}_{i=1}^N$ denotes the collection of computational and networking nodes (\eg routers, Cloud servers, IoT devices), with resources specified by the asset bucket of node capabilities $\noderes$ (\eg computation power, memory, storage, availability);
   \item $\links = \{(i, j, \bass_{ij})\}_{i,j \in N}$ describes a collection of end-to-end links that interconnect nodes in $\nodes$, with resources specified by the asset bucket of link capabilities $\linkres$ (\eg physical bandwidth, latency).
    Furthermore, we define $\pathres$ as the bucket of path-level capabilities, which extends $\linkres$ by aggregating link capacities across paths (\ie an ordered sequence of links). Note that the featured bandwidth is additive on a link and convex over a path.
\end{itemize}

\subsubsection{Application}\label{def:appl}The application $\app$ is defined as a 5-tuple $(\services, \interactions, \servicereq, \intreq, \flows)$, where:
\begin{itemize}
    \item $\services = \{(i, \bass_i)\}_{i=1}^M$ denotes a set of $M$ services (\eg micro-services, databases, Virtual Network Functions), each presenting the requirements $\bass_i$ defined according to the requirement specification $\servicereq$, compatible to $\noderes$ in the infrastructure model;
    \item $\interactions = \{(i, j, \bass_{ij})\}_{i,j \in M}$ denotes the set of interactions among application services, each with its requirement bucket $\bass_{ij}$ defined according to the requirement specification $\intreq$;
    \item $\flows$
    is the set of flows, \ie ordered lists of application services in the form $[s_1, s_2, \ldots]$, dictating the critical paths of data exchanges, where a critical path is a series of service interactions that need to be executed in a designated sequence to ensure the correct and timely outcome of the application's operation.
\end{itemize}
In the application model, the domains of requirement specification $\servicereq$ and $\intreq$ must correspond to the domains of capability specification $\noderes$ and $\linkres$ in the infrastructure model, ensuring consistency in the overall environment model. 
In the context of the infrastructure nodes, the function $\sat$ is used to signify the quality level of an asset value. Conversely, in application settings, it represents the strictness of a requirement (\eg requesting 128 CPU cores is \textit{more stringent} than requesting only 16).

\subsubsection{Placement}\label{def:placement}Let $\app^i$ be the $i$-th multi-service application with requirements ($\servicereq, \intreq$) to be deployed on infrastructure  $\infra$ with node and path resources ($\noderes, \pathres$). We define the placement as the function $\placement^i: \services^i \rightarrow \nodes$ that maps each service of the application $\app^i$ to a node of the infrastructure $\infra$.
Let $\placement^i(s_j) = n$ represent the placement of the $j$-th service of application $\app^i$ on node $n$ within infrastructure $\infra$. Since each node $n \in \nodes$ can host services from different applications, we define $\barplacement(n)$ as the function that returns the collection of services assigned to node $n$. 
The set of placement functions is combinatorial in the number of infrastructure nodes and the number of services of the given applications~\cite{FogTorchPI}. However, the space of valid placements is constrained by the application's requirements, the infrastructure's capabilities, and the service interactions.
A placement is \textit{valid} when it satisfies the following constraints:
\begin{itemize}
    \item the aggregated resource requirements of all services placed on a node must meet the node's resource capacities, \ie for each node $n \in \nodes$, the following constraint must hold:
    \begin{equation*}
        \bigoplus_{s \in \barplacement(n)} \bass_s \prec \hat{\bass}_n
    \end{equation*}
    where the aggregation is performed according to the $\noderes$ bucket, and $\hat{\bass}_n$ is the residual resource capacity of $n$;
    
    \item for each interaction in the application $\app^i$ involving source $s_j$ and destination $s_k$, there exists a path between the nodes $\placement^i(s_j)$ and $\placement^i(s_k)$ whose residual resources meet the interaction requirements. 
\end{itemize}
It is important to note that these constraints apply to the \textit{residual} resources of the infrastructure, meaning resources not yet allocated to the services deployed on the nodes and the interactions across the paths. The residual resources are determined by subtracting the total resource requirements of the services and interactions from the resource capacities of the nodes and paths, respectively.

%% file: fig/environment.tex
\begin{figure}
    \centering
    \includegraphics[width=.9\linewidth, trim={18.5cm 1.2cm 18.5cm 1.2cm}, clip]{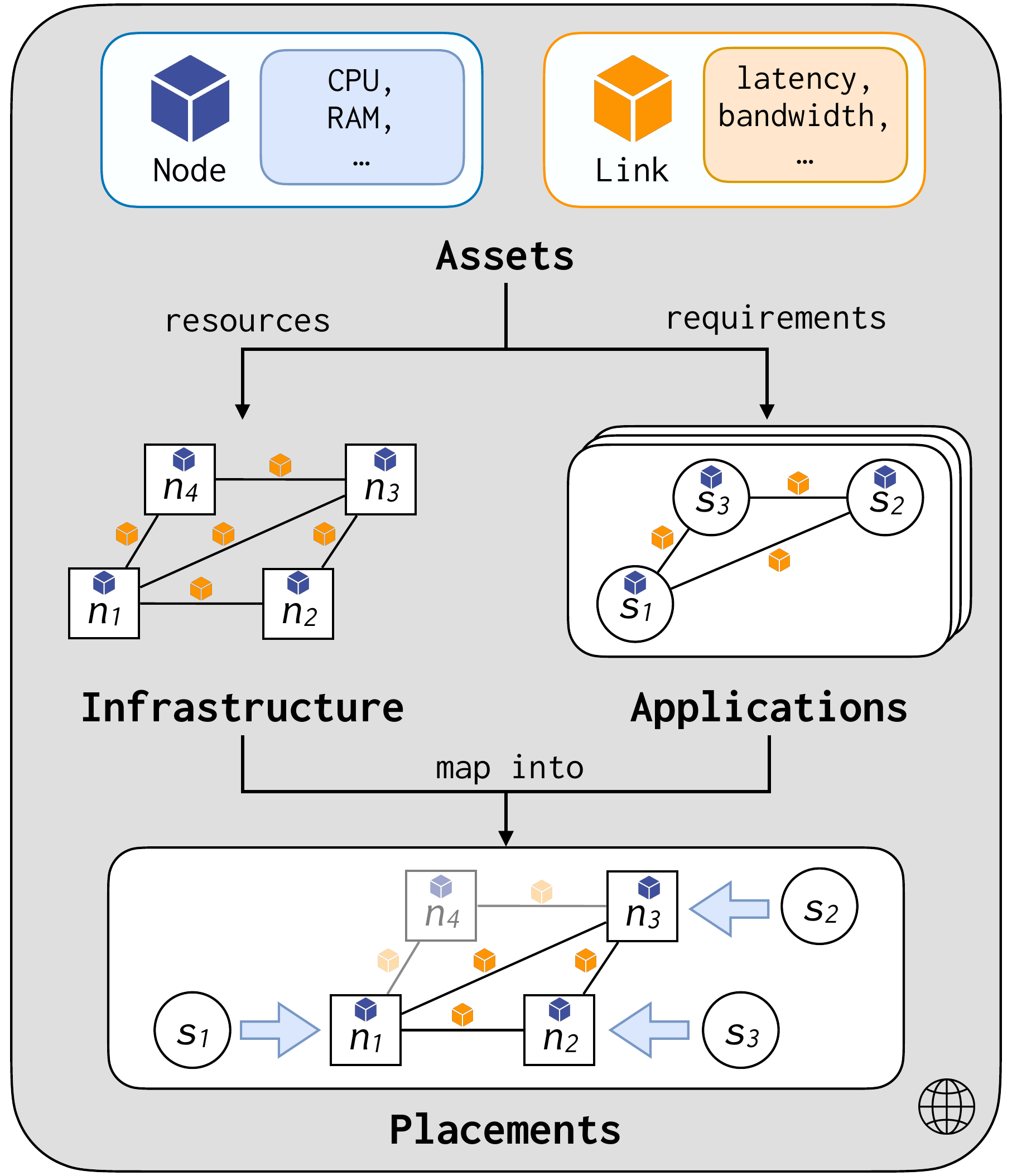}
    \caption{Representation of the \eclypse environment. \fncd{Assets} characterise entities and relations as \fncd{Infrastructure} resources and \fncd{Application} requirements (\eg number of CPUs, latency\dots). \cdp{Placement} map application services to infrastructure nodes.}
    \label{fig:environment}
\end{figure}

%% file: tab/notation.tex
\begin{table}[t]
\footnotesize
    \centering
    \caption{Notation of the Environment model}
    \label{tab:notation}
    \begin{tabular}{l|l|l}
    \toprule
    \textbf{Asset ($\asset$)} & \textbf{Infrastructure ($\infra$)} & \textbf{Application ($\app$)} \\
    \midrule
    $\ass$: Asset value & $\nodes$: Set of nodes & $\services$: Set of services \\
    $\domain$: Domain of $\ass$ & $\links$: Set of links & $\interactions$: Set of interactions \\
    $\sat$: Comparison function & $\noderes$: Node assets & $\servicereq$: Service assets \\
    $\agg$: Aggregation function & $\linkres$: Link assets & $\intreq$: Interaction assets \\
    $\assets$: Asset bucket & $\pathres$: Path assets& $\flows$: Set of flows\\
    $\bass$: Asset bucket values & & \\
    \bottomrule
    \end{tabular}
\end{table}

%% file: tab/assets.tex
\begin{table*}[!ht]
\renewcommand{\arraystretch}{1.2} 
\caption{Asset types classification.}
\centering
\begin{tabular}{|l|l|l|l|l|}
\hline
\textbf{Asset type} & \textbf{Domain} & \textbf{Comparison ($\sat$)} & \textbf{Aggregation ($\agg$)} & \textbf{Purpose (Example)} \\ \hline
\textbf{Additive} & $\{\ass \in \posreals \mid \lb \leq \ass \leq \ub \}$ & $\ass_1 \leq \ass_2$ & $\ass_1 + \ass_2$ & Cumulative \textit{total} resource (CPU). \\ \hline
\textbf{Concave} & $\{\ass \in \posreals \mid \lb \leq \ass \leq \ub \}$ & $\ass_1 \leq \ass_2$ & $\max(\ass_1, \ass_2)$ & Unique \textit{strongest} resource (link latency). \\ \hline
\textbf{Convex} & $\{\ass \in \posreals \mid \lb \geq \ass \geq \ub \}$ & $\ass_1 \geq \ass_2$ & $\min(\ass_1, \ass_2)$ & Unique \textit{weakest} resource (path bandwidth). \\ \hline
\textbf{Multiplicative} & $\{ \ass \in [0,1] \mid \lb \leq \ass \leq \ub \}$ & $\ass_1 \leq \ass_2$ & $\ass_1 \times \ass_2$ & \textit{Combined} effect of resources (reliability). \\ \hline
\textbf{Symbolic} & $\{\ass \mid \ass \cap \lb = \lb,\, \ass \cup \ub = \ub;\, \ass, \lb, \ub \subseteq \mathcal{U}\}$ & $\ass_1 \subseteq \ass_2$ & $\ass_1 \cap \ass_2$ & \textit{Categorical} values (access levels). \\ \hline
\end{tabular}
\label{tab:assets}
\end{table*}

%% file: src/model/simulation.tex
\input{fig/simulation}

\subsection{Simulation Model}
\label{sec:model_sim}
The environment model describes \textit{a} state of all the entities describe in ~\cref{sec:model_env}, at a \textit{specific} moment in time (\ie a \textit{snapshot}). It is thus crucial to define how the environment evolves, along with an appropriate set of functionalities to access and evaluate its state. Cloud-Edge environments are characterised by heterogeneous events that may continuously alter the environments' state. Building and reproducing simulations that accurately catch such dynamicity is challenging. To address this, we provide a flexible framework to define arbitrarily complex simulations.
    
\subsubsection{Simulation Graph}\label{sec:sim-graph}
As graphically shown in the left-hand side of \Cref{fig:simulation}, our framework is based on the notion of \textit{simulation graph}, a computational directed acyclic graph (DAG) where nodes represent operational steps in the simulation, while edges represent the ``\cd{activates}'' relation according to their direction (\ie the completion of the operation by the source node \cd{activates} the operation in the destination node).

We distinguish three categories of nodes, depending on their functional role within the simulation process:
\begin{itemize}
    \item a \textit{trigger} $\trg \in \Trg$ is a node that \textit{originates} a sequence of operations in the simulation graph. It does not encapsulate any operational logic and serves purely as an activation source. Triggers may be fired either \textit{manually} (\eg start the simulation) or \textit{programmatically}, based on temporal (\eg scheduled ticks) or cascading rules (\eg in response to an event);
    
    \item an \textit{event} $\evt \in \Evt$ is a node that \textit{applies} a transformation to the simulation state, such as modifying resources or enforcing placements. Events can be activated by one or more triggers $\trg \in \Trg$;
    
    \item a \textit{callback} $\cbk \in \Evt$ is a special type of event whose role is to \textit{observe} and \textit{analyse} the state immediately after another event completes. Callbacks do not alter the environment but extract information from it (\eg compute metrics or collect logs). Each callback is activated directly by a single event and is executed \textit{immediately} after its completion.
\end{itemize}

This conceptual separation clarifies the semantic role of each node type in the simulation workflow, distinguishing state-changing transitions (events) from monitoring operations (callbacks), while preserving a unified model of event-driven execution governed by activation edges (triggers).

\subsubsection{Simulation Workflow}\label{sec:sim-workflow}
As per the graph's structure in \cref{fig:simulation}, there is no limitation regarding the out-degree of each node in the simulation graph (\ie any trigger may activate an arbitrary number of nodes). Once a node's operation is completed, the execution of all its child nodes can run in parallel. Although this might seem beneficial from a computational standpoint, it might undermine the goal of maintaining a controlled and reproducible simulation. For this reason, we define the \textit{operational workflow} as the ordered sequence of events induced by the depth-first traversal rooted on a trigger $t$. An example of the resulting sequence of operations is depicted in the right-hand side of \cref{fig:simulation}:
\begin{enumerate*}[label=\textit{(\roman{*})}]
    \item the manual trigger activates $\evt_1$, and $\cbk_1$ reads the produced state right after its completion;
    \item $\evt_2$ and 
    \item $\evt_3$ are executed sequentially. Finally
    \item $\evt_4$ is executed, activating $\cbk_1$ and $\cbk_2$ to read and process the environment's state.
\end{enumerate*}
%

We define three \textit{default events} that serve as the essential building blocks for simulations:

\begin{enumerate}
    \item \textit{Update policy}: this function maintains environmental dynamism over time by adjusting both the infrastructure's resource capabilities and the application's requirements as needed. The seamless integration of this event's definition and its activation -- either through specific or periodic triggers -- facilitates the creation of complex simulations.

    \item \textit{Placement lookup}: this event allows for finding the placements (\cref{def:placement}) for each application based on a defined \textit{placement strategy}, which implements the search algorithm to select appropriate placements. Its flexibility supports simulations that analyse infrastructure and application behaviour under varying placement strategies.

    \item \textit{Placement fulfilment}: this ensures that the current infrastructure status can satisfy the requirements of the placed applications. It involves enforcing placement constraint satisfaction by allocating resources to valid placements (as described in \cref{def:placement}) and resetting those that cannot be met.
\end{enumerate}
The state produced by executing these three events accounts for the allocated resources and will be the input for subsequent events.
All events are scheduled and executed using an asynchronous programming model: although the Python Global Interpreter Lock (GIL) prevents actual concurrent execution, the simulation engine ensures non-blocking, sequentially consistent execution of events within an event loop. User-defined callbacks -- such as those for metric collection or report generation -- are likewise registered and executed asynchronously, ensuring that other events do not interfere with the simulation's logical flow.

%% file: fig/simulation.tex
\begin{figure*}[ht!]
    \centering
    \includegraphics[width=.8\linewidth, trim={0cm 6cm 0cm 6cm}, clip]{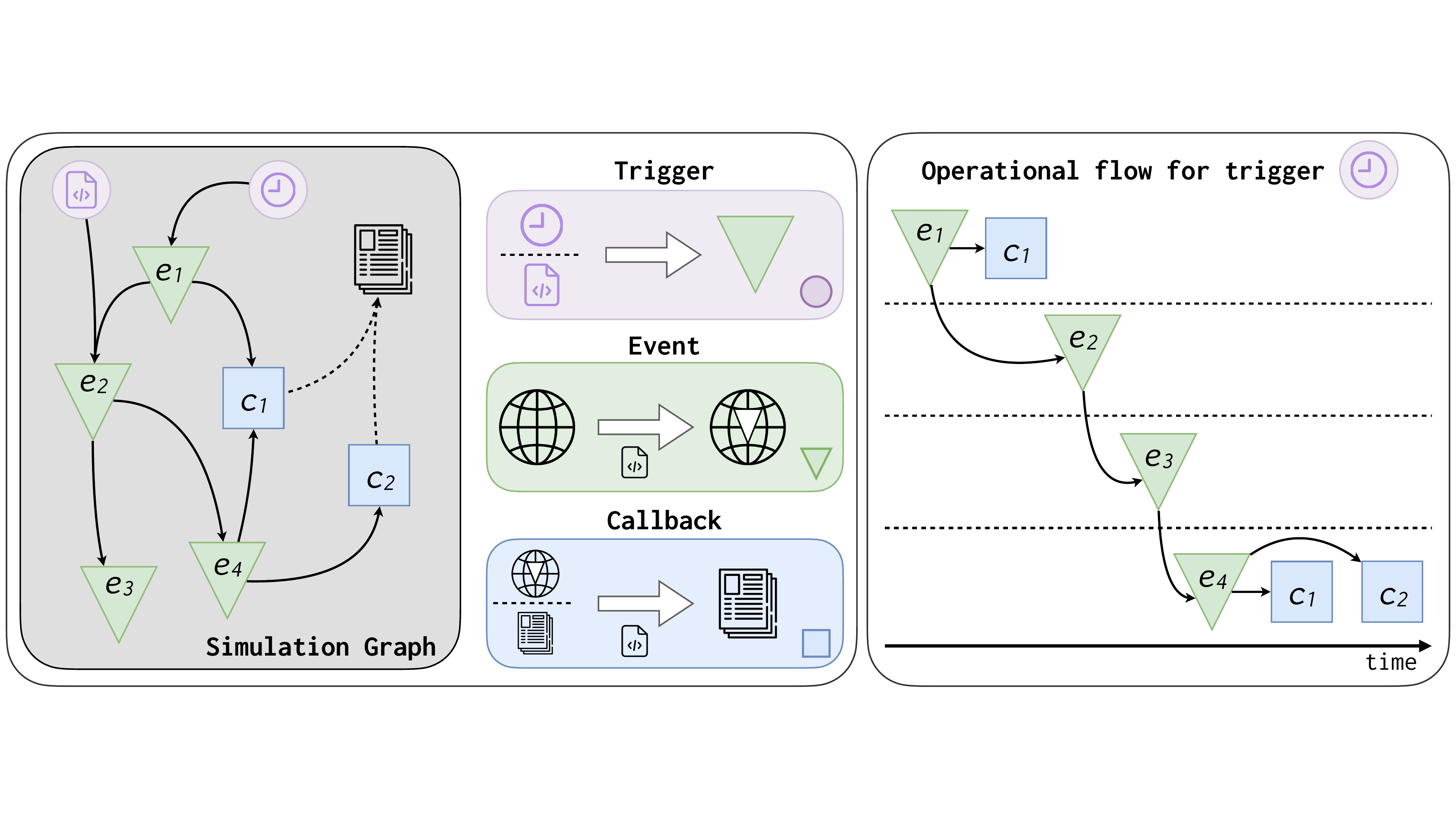}
    \caption{Representation of the simulation model of \eclypse. The left-hand side shows the simulation graphs, where purple nodes represent triggers, green nodes represent events, blue nodes represent callbacks, and arrows model the \texttt{activates} relationship. The right-hand side defines how a trigger originates an operational flow, events modify the state of the simulation, and callbacks read and process the state \textit{right after} the completion of the activating event.}
    \label{fig:simulation}
\end{figure*}

%% file: src/framework/main.tex
\section{Simulation \& Emulation}
\label{sec:framework}

\input{src/framework/architecture.tex}
\input{src/framework/simulation.tex}
\input{src/framework/emulation.tex}

%% file: src/framework/architecture.tex
The \eclypse framework's software stack consists of two main layers, as sketched in~\cref{fig:stack}.

\input{fig/stack}

We designed the core layer to ease the use of the framework by abstracting the intricate internal mechanics of \eclypse. To achieve this, the core integrates two key external dependencies: NetworkX\footnote{\url{https://networkx.org}} for efficient graph handling and Ray\footnote{\url{https://ray.io}} for seamless parallelisation and scaling of distributed processes. 
These tools underpin the robust execution of simulations and emulations. The entire bottom is implemented in Cython\footnote{\url{https://cython.org}}. Pre-compiled packages are provided for a range of standard processors and operating systems, ensuring broad compatibility and ease of deployment.

By wrapping the core layer, \eclypse features a highly modular and extensible architecture, empowering users to rapidly simulate and emulate Cloud-Edge environments while focusing solely on the aspects most relevant to their needs. This modularity is realised via a class structure corresponding to the mathematical model described in \cref{sec:model}, facilitating a smooth shift from theoretical formulation to practical application.
We organised the architecture into distinct packages, each addressing specific facets of the simulation process and mirroring the structured approach of the underlying model:
\begin{itemize} 
    \item \textit{Environment Model}: represented by the \cd{graph} and \cd{placement} packages, which provide tools for defining and managing infrastructures and applications, including off-the-shelf or custom placement strategies. 
    \item \textit{Simulation Model}: encapsulated in the \cd{workflow} package, allows implementing events and callbacks, enabling users to build arbitrarily complex simulation graphs. 
    \item \textit{Transition to Emulation}: addressed by the \cd{remote} package, which provides functionalities to implement actual services within the applications. 
\end{itemize}
Each package comes with pre-configured tools that enable rapid environment setup, and the distinct allocation of roles among packages aids in the seamless expansion of the framework's functionalities.

%% file: fig/stack.tex
\begin{figure}
    \centering
    \includegraphics[width=.9\linewidth, trim={0cm 3cm 0cm 0cm}, clip]{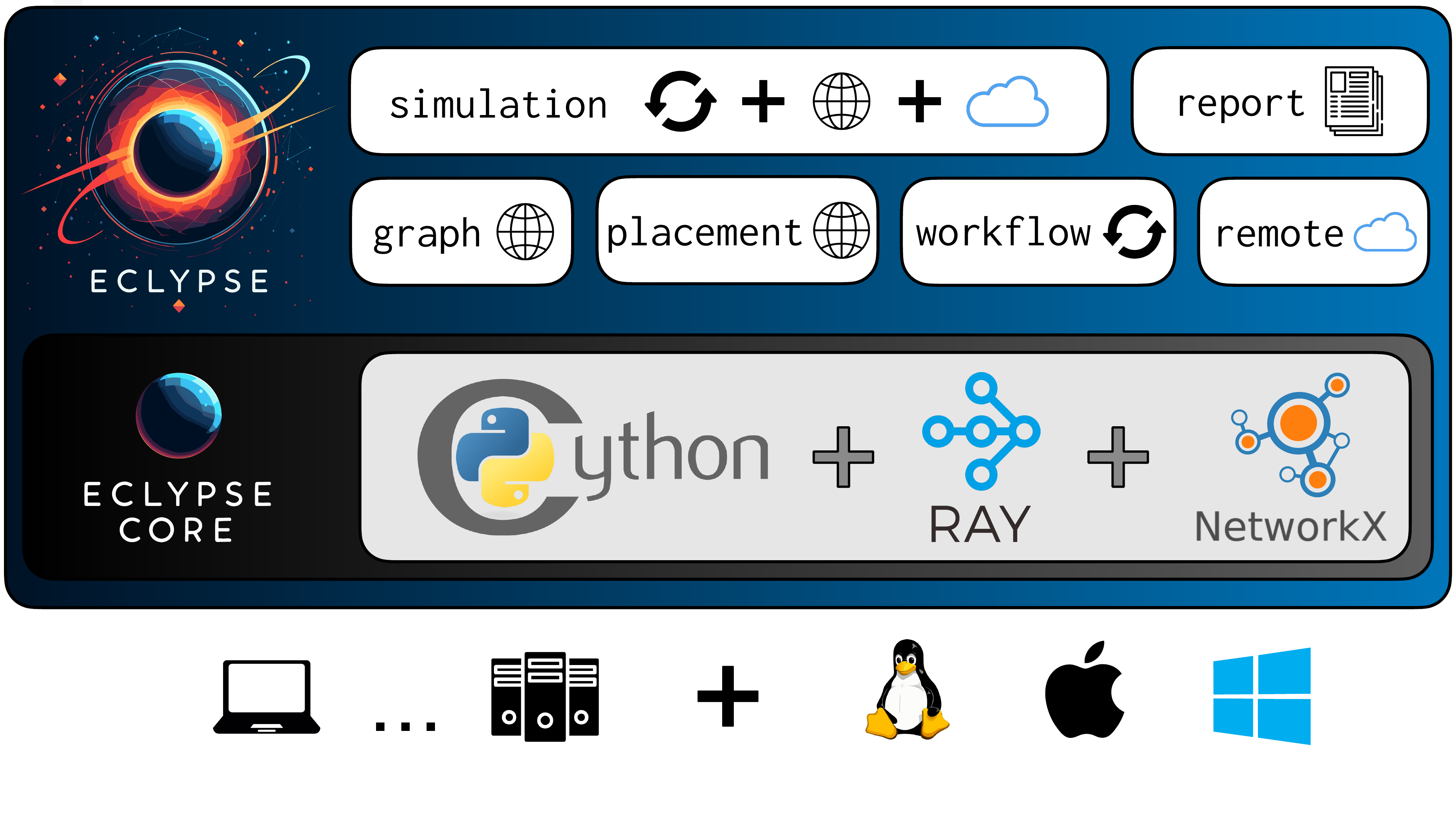}
    \caption{Software stack of \eclypse.}
    \label{fig:stack}
\end{figure}

%% file: src/framework/simulation.tex
\subsection{Simulation}
\label{sec:framework_simulation}
The core principle behind \eclypses simulation functionalities is to offer tools for fine-grained, user-friendly control, balancing realism, computational efficiency, and minimal coding effort. The simulation workflow involves five key steps, depicted in~\cref{fig:workflow}. Namely: 
\begin{enumerate*}[label={}]
    \item asset definition (\cref{sec:simframe-asset-definition}), 
    \item environment building (\cref{sec:simframe-env-building}),
    \item simulation planning (\cref{sec:simframe-sim-planning}),
    \item execution (\cref{sec:simframe-execution}), and
    \item reporting (\cref{sec:simframe-reporting}).
\end{enumerate*}

\subsubsection{Assets definition}\label{sec:simframe-asset-definition}
All the elements participating in the simulation environment are characterised by their \cd{Assets} (\cref{def:asset}), whose definition represents the first step to setting up a simulation.
To this end, the \cd{Asset} class (from the \cd{graph} package) provides the abstraction to define the asset through its domain, the aggregation function as the \cd{aggregate} method, and the comparison function as the \cd{compare} method. Furthermore, we provide five subclasses implementing the methods above to match the behaviours of assets in \cref{tab:assets}. 

Furthermore, we specialised these classes to offer a collection of default assets that cover the standard characterisation of nodes and links. Default assets for nodes include \textit{CPU}, \textit{RAM}, \textit{storage}, \textit{GPU}, \textit{availability}, and \textit{processing time}. \textit{Latency} and \textit{bandwidth} represent the default assets for links, and are the basis for modelling network performance.

\subsubsection{Environment building}\label{sec:simframe-env-building}
The two main entities which take part in the simulation are the infrastructure and the applications, implemented through the \cd{Infrastructure} (\cref{def:infra}) and \cd{Application} (\cref{def:appl}) classes, respectively.
These classes extend the \cd{AssetGraph} class, allowing asset management to be incorporated within any NetworkX graph. Building an infrastructure consists of initialising the class, with the set of assets representing the resources, the definition of a placement strategy to provide a default behaviour in placing applications, and the update policy to implement the dynamics of the resources over time.
After initialisation, nodes and edges are introduced using methods that ensure consistency with the respective node and asset
(\eg Cloud nodes and their networks feature more resources compared to IoT nodes). The same approach holds for applications, whose initialisation necessitates the specification of flows to estimate response time at runtime. \eclypse provides predefined builders to ease the creation of infrastructures, such as \cd{hierarchical} for tiered network structures, \cd{star} for centralised network topologies, \cd{random} for network topologies with a specific degree of randomness.

\input{fig/workflow}

\subsubsection{Simulation planning}\label{sec:simframe-sim-planning}
This second step defines the \textit{dynamics} of the simulation by specifying the set of operational nodes -- \ie instances of \cd{EclypseEvent} -- and the associated triggers regulating their activation and execution flow.

Each \cd{EclypseEvent} encapsulates a self-contained computational unit that can read or modify the simulation state. Its core logic is implemented via the \cd{\_\_call\_\_} method, whose input signature is automatically determined by the event's \cd{type} attribute. Such attribute specifies the abstraction level at which the event operates -- \ie simulation, infrastructure, application, node, service, link, or interaction level -- and ensures that the event receives the appropriate slice of the internal state to perform its task.

Events can optionally be flagged as \cd{is\_callback=True}, in which case they behave as reactive post-processing blocks. These are automatically executed immediately after the event that triggered them and are not scheduled independently. This allows users to define metric extraction, reporting, or logging logic that remains coupled to upstream events.

The execution logic is orchestrated through a flexible trigger system. Each event must be associated with one or more triggers, which determine when and under what conditions it is executed. Triggers can be defined either by subclassing the base \cd{Trigger} class or by using the provided off-the-shelf implementations. Available trigger types include \textit{periodic}, \textit{scheduled} and \textit{random} triggers, all available in both time-based and cascade-based variants. Cascade triggers support dependency-based execution, enabling complex workflows where events activate subsequent events in response to simulation state changes.

The simulation workflow is built by defining and registering the events, each paired with its trigger logic. After this, the simulation can be \textit{compiled}, thus the framework analyses the event graph, resolves trigger chains and generates the corresponding event-driven execution plan.

\subsubsection{Execution}\label{sec:simframe-execution}
To run a simulation, it is sufficient to assemble everything by instantiating the \cd{Simulation} class. All simulation settings are encapsulated in a \cd{SimulationConfig}, which, in addition to the details discussed in \cref{sec:simframe-sim-planning}, determines the duration of the simulation and its granularity (\ie the frequency of a base periodic \texttt{step} event). For example, a simulation might run for 24 hours to capture peak and off-peak behaviours or to simulate specific failure scenarios in the infrastructure.

By design, a newly instantiated simulation is empty: it contains no predefined assets, applications, triggers, events or metrics. While \eclypse offers a set of off-the-shelf components -- such as assets, placement strategies, metrics and reporting utilities -- none of them are automatically attached to the simulation. This ensures full user control and guarantees that every simulation is explicitly defined according to its intended scope.

The only predefined elements are the simulation’s core events (\ie \cd{start}, \cd{stop}, \dots) which provide the minimal structure required to initialise, advance, and terminate a simulation. These events serve as structural anchors but do not impose any behaviour beyond their registration. Notably, their functionality is fully customisable: users may override or extend their logic by redefining the corresponding event handlers, allowing even the simulation core to be adapted to specific experimental needs.

Once the simulation is started using the \cd{start} method, the simulator compiles the simulation graph to identify triggers and ensures that the structure of the nodes and the \cd{activates} relations follow the constraints specified in \cref{sec:sim-graph}. The simulator then handles the registered events in two main ways, according to the compiled triggers: it responds to manually triggered events by scheduling them for execution, and activates periodic events according to the simulation clock.

\subsubsection{Reporting}\label{sec:simframe-reporting}
Callbacks, defined as \textit{reportable} events, enable the extraction and storage of simulation insights in various formats. By specifying a reporting mode during their configuration, users can persist the output of a callback in formats such as \cd{csv} for tabular data, \cd{json} for hierarchical structures, \cd{tboard} for TensorBoard\footnote{\url{https://www.tensorflow.org/tensorboard}} integration, or \cd{gml} for graph-based visualisations.

The scope and structure of the reported data are determined by the callback's \cd{type}, which identifies the target entity from which the simulation state is read (\eg application, infrastructure, node, interaction), ensuring that each report focuses precisely on the desired simulation layer.

To simplify reporting setup, \eclypse provides a collection of off-the-shelf callbacks such as the monitoring all predefined assets, service placements, application response times based on given flows, the number of alive nodes and the overall simulation time.

%% file: fig/workflow.tex
\begin{figure}
    \centering
    \includegraphics[width=0.9\linewidth, trim={20cm 0cm 20cm 0cm}, clip]{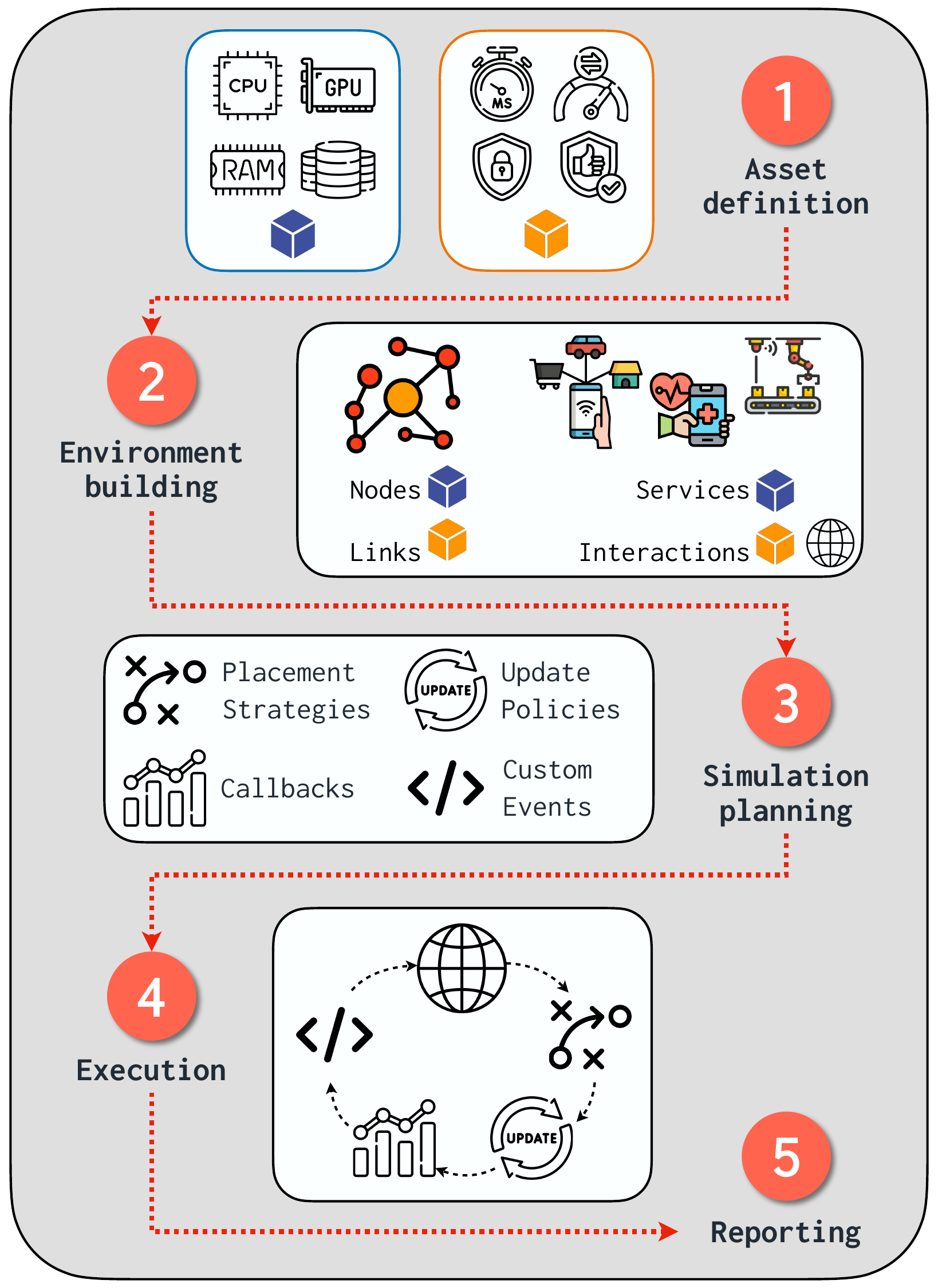}
    \caption{Five steps of the \eclypse simulation workflow. The figure summarises the end-to-end process required to configure and execute a simulation within \eclypse: \textit{(1) Asset definition}, where nodes and links are characterised by default or custom properties; \textit{(2) Environment building}, in which the infrastructure and application graphs are instantiated with placement and update policies; \textit{(3) Simulation planning}, where events and callbacks are defined and linked to form the operational flow; \textit{(4) Execution}, which runs the simulation through asynchronous, sequentially consistent scheduling; and \textit{(5) Reporting}, which collects and formats results via structured, non-intrusive callbacks.}
    \label{fig:workflow}
\end{figure}

%% file: src/framework/emulation.tex
\subsection{Emulation}
\label{sec:framework_emulation}

The emulation functionalities of \eclypse aim to showcase and validate the behaviour of real application implementations within a controlled, virtualised infrastructure. While conceptually rooted in the simulation model (\cref{sec:framework_simulation}), the emulation workflow enables the deployment and execution of actual Python services that communicate across an emulated environment. These capabilities are powered by the Ray library, which supports distributed task execution through its actor model. As depicted in~\cref{fig:emulation}, the emulation process is structured in four key steps.
Namely:
\begin{enumerate*}[label={}]
    \item remote configuration (\cref{sec:emuframe-config}), 
    \item remote environment building (\cref{sec:emuframe-env-building}),
    \item emulation planning (\cref{sec:emuframe-emu-planning}) and
    \item execution (\cref{sec:emuframe-execution}).
\end{enumerate*}

\input{fig/emulation}

\subsubsection{Remote configuration}
\label{sec:emuframe-config}
In this initial step, the user specifies the computational resources assigned to each infrastructure node, which will be emulated as Ray actors. This configuration is enabled by setting the \cd{remote=True} flag when defining the \cd{Simulation} object. Additionally, the user defines the communication interface type to be used between services (\eg REST or MPI). Optionally, Ray-specific options such as resource limits and scheduling policies can be assigned to each actor using the \cd{RemoteBootstrap} class.

\subsubsection{Remote Environment building}
\label{sec:emuframe-env-building}
Once the infrastructure configuration is complete, \eclypse deploys a Ray cluster and instantiates remote actors for each infrastructure node, depending on the \cd{RemoteBootstrap} instance defined at configuration time (\eg the policy to dispatch actors across the cluster). The user must also provide the implementation of each service using the \cd{add\_service} method within the \cd{Application} definition. Services are defined as subclasses of the \cd{Service} class and encapsulate business logic, such as data processing or communication handling. Interactions among services are specified based on the chosen communication interface, which implicitly govern how messages are encoded, routed, and delivered during emulation.

\subsubsection{Emulation planning}
\label{sec:emuframe-emu-planning}
At this stage, the user defines message flows and may specify remote events by using the \cd{remote=True} flag in their constructor or decorator. These events are executed directly on remote services or nodes and are used to monitor or influence the real application behaviour. Then, the simulator gathers the results and processes them as per standard events and callbacks (viz. \cref{sec:simframe-sim-planning}).
Thus, the same planning logic can be reused across both simulation and emulation setups, with minimal changes. In most cases, enabling emulation requires only activating the remote execution flag, without restructuring callback or event logic.

\subsubsection{Execution \& Runtime}
\label{sec:emuframe-execution}
When the simulation begins, the placement process identifies suitable nodes for each application. Once placement is found, the corresponding service instances are deployed on their designated Ray actors, and their internal state is prepared via the \cd{on\_deploy} method. If a placement becomes invalid, the \cd{on\_undeploy} hook is triggered, and the services are halted and undeployed.
During execution, services interact using the \cd{EclypseCommunicationInterface}, which abstracts network protocols and models their performance impact. Currently supported interfaces include:
\textit{REST}, for HTTP-based request/response communication typical of web applications, and \textit{MPI}, for message-passing patterns in HPC and parallel environments.

A typical service interaction follows the process illustrated in step (4) of~\cref{fig:emulation}. For instance, when a service $s_1$ sends a message to $s_2$, the following sequence occurs:
\begin{enumerate}
    \item $s_1$ passes the message to its hosting node $n_1$;
    \item $n_1$ requests a communication route to the simulator;
    \item the simulator computes a route to $n_2$ (host of $s_2$) and returns the associated cost;
    \item $n_1$ transmits the message to $n_2$, applying the delay based on route cost\footnote{Route cost is computed based on link latency and message transmission time, which depends on bandwidth and message size.};
    \item $n_2$ delivers the message to $s_2$, via suitable demultiplexing.
\end{enumerate}

%% file: fig/emulation.tex
\begin{figure}[t]
    \centering
    \includegraphics[width=0.9\linewidth, trim={20cm 0cm 20cm 0cm}, clip]{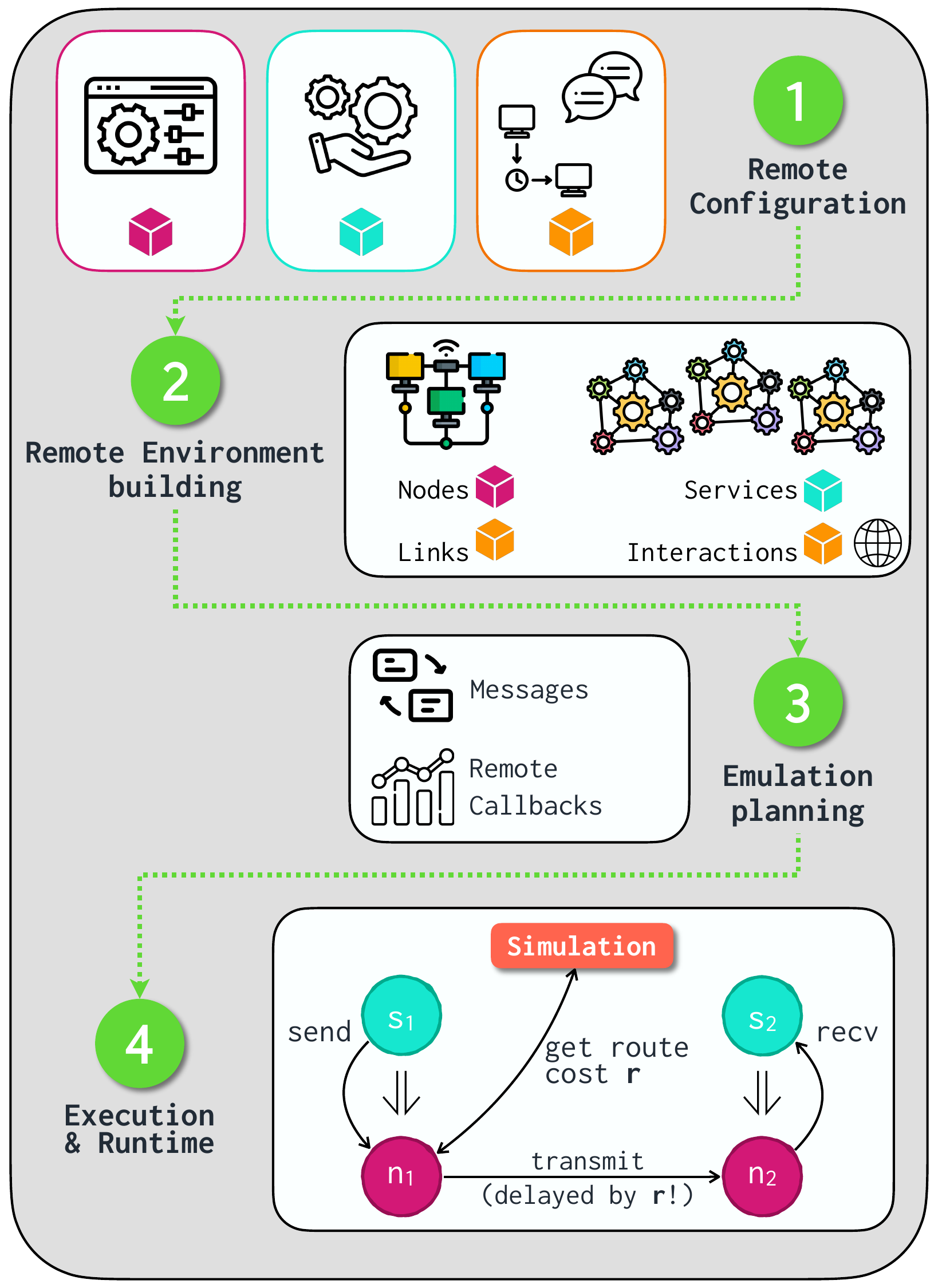}
    \caption{Overview of the emulation workflow in \eclypse. The process is structured in four main steps: 
    \textit{(1)} \textit{Remote configuration}, which defines the actual resource allocation for each remote node and the communication interface types used between deployed services;
    \textit{(2)} \textit{Remote environment building}, consisting of the deployment of a Ray cluster, the definition of services' logic, and the setup of service interactions based on the specified communication interfaces;
    \textit{(3)} \textit{Emulation planning} through the definition of message flows and remote callbacks for monitoring and coordination;
    \textit{(4)} \textit{Execution}, where emulated services interact over the virtualised infrastructure. The lower part illustrates a message exchange between two services $s_1$ and $s_2$ deployed on nodes $n_1$ and $n_2$, with communication subject to a simulated delay cost $r$.}
    \label{fig:emulation}
\end{figure}

%% file: src/experiment/main.tex
\section{Use Cases}
\label{sec:experiments}

This section describes three distinct use cases designed to test\footnote{Experiments were run on a machine featuring Ubuntu 22.04.5 LTS (GNU/Linux 5.15.0-1065-nvidia x86\_64) equipped with 1.5TB of RAM and two 3.7GHz AMD EPYC 9534 64-Core processors, Python 3.11} \eclypses flexibility and effectiveness in various infrastructure and application scenarios, exploring different aspects of Cloud-Edge systems\footnote{Results' analysis is available at: \url{https://github.com/eclypse-org/eclypse/tree/examples-analysis}}:
\begin{enumerate}
    \item A \textit{grid analysis} that evaluates various application placement strategies across multiple network topologies and sizes, and resource availability conditions (\cref{sec:usecase1}).
    \item A \textit{user distribution} analysis based on a real-world dataset which examines how different user workloads affect infrastructure and application performance (\cref{sec:usecase2}).
    \item A \textit{remote image prediction} experiment simulating the performance of an AI-based application for distributed image processing, demonstrating how \eclypse can emulate AI workloads across Edge/Cloud environments (\cref{sec:usecase3}).
\end{enumerate}

The stakeholders for each use case vary according to the objectives addressed. For \textit{UC1}, the stakeholders may be researchers and infrastructure providers that aim to optimise resource utilisation under various conditions. \textit{UC2} focuses on application developers seeking to enhance user experience during fluctuating workloads. Lastly, \textit{UC3} targets AI application developers and/or system engineers interested in testing and refining machine learning models in distributed environments.

For each scenario, we describe the key components of the simulation environment (\cref{sec:model_env}), thus assets, applications, and infrastructure. 
The scenarios share most of the simulation workflow, which is defined by a set of events $\evt \in \Evt$ and callbacks $\cbk \in \Cbk$, including the default events described in the remainder of \cref{sec:model_sim} and the off-the-shelf metrics listed in~\cref{sec:simframe-reporting}, respectively. Additionally, we added context-aware events and callbacks to each use case, showcasing \eclypse's simplicity and modularity.

Finally, we monitor and analyse the CPU and memory usage of \eclypse itself by implementing three simulation metrics (as callbacks) that leverage the \textit{psutil} Python library\footnote{\url{https://pypi.org/project/psutil/}}, with CPU usage measured as a percentage of the single core assigned to the simulator process. Monitoring these metrics for UC1 (\cref{sec:usecase1}) was not feasible since we used Ray Tune, which automatically manages all physical resources, impairing accurate monitoring of the overall resource consumption.

\input{src/experiment/grid-analysis}
\input{src/experiment/user-distribution}
\input{src/experiment/image-prediction}

%% file: src/experiment/grid-analysis.tex
\begin{figure*}[t]
    \centering
    \includegraphics[width=.99\textwidth]{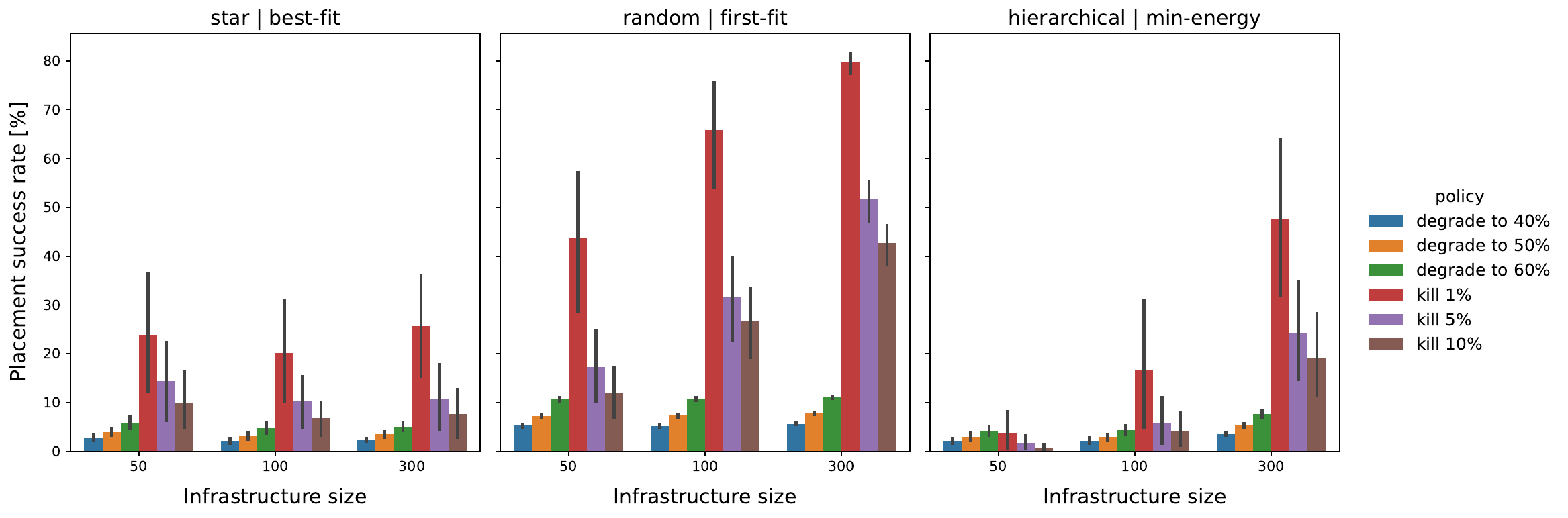}
    \caption{Placement success rate in UC1. We report the best-performing placement strategy for each topology at varying sizes and updated policies.}
    \label{fig:pl_success_rate}
\end{figure*}

\subsection{Placement strategies comparison}
\label{sec:usecase1}

In the first use case, an infrastructure provider aims at optimising the use of resources across its data centres and edge nodes, employing the \eclypse framework to evaluate different application placement strategies. The provider manages various applications from various sectors, including healthcare, Industry 4.0, and e-commerce. The evaluation aims at determining which placement strategy best balances resource efficiency and application performance under different network conditions. To conduct the experiments, we selected three distinct applications: CPuS-IoT~\cite{cpusiot}, a cyber-physical IoT framework for manufacturing systems; Health Guardian~\cite{healthguardian}, a digital health monitoring application; and VAS~\cite{vas}, a Video Analytics Serving application.

\subsubsection{Simulation environment}
The environment for this scenario includes an infrastructure alongside the three applications. This setup is tested under varying conditions that simulate different network topologies (star, hierarchical, and random), and network sizes (50, 100, and 300 nodes). Moreover, the initial infrastructure load is varied to simulate different saturation levels for each asset(0\%, 25\%, 50\%, and 75\%), allowing for a comprehensive analysis of the system's behaviour under different resource availability constraints. Each simulation runs for 600 ticks, and for each configuration, three random seeds are employed to ensure the robustness of the results.

\subsubsection{Simulation Workflow}
We executed three strategies for the placement lookup (denoting a default event in $\Evt$): the \cd{first-fit} strategy places applications on the first available node that meets the resource requirements; the \cd{best-fit} strategy, on the other hand, selects the node that maximises resource utilisation without exceeding capacity. Finally, the \cd{min-energy} strategy prioritises nodes with the lowest energy consumption, where the energy consumed by each node is computed as a linear combination of the resources allocated to it (\ie CPU, GPU, RAM, and storage).

In addition to placement strategies, we defined two update policies to apply runtime changes to the network, thus simulating real-world dynamics. The policy \cd{degrade(X)} gradually reduces the available node resources from 100\% to a specified value $X\%$, simulating continuous resource degradation over time. The policy \cd{kill(X)} deactivates the nodes with a probability of $X\%$ and reactivates them with a probability of $\frac{X}{2}\%$, capturing the random failure and recovery of the nodes.
Each simulation includes one of the two policies implemented within the events $\Evt$. We enriched the callback set $\Cbk$ to collect the placement success rate of each application.

\input{tab/grid-params}

\subsubsection{Results}
Considering all the varying parameters (summarised in \cref{tab:grid_params}) and their combinations, we performed 3024 simulations lasting 600 ticks, with an average simulation time of 240 seconds (4 minutes) and 2.5 ticks/s.
We leveraged Ray Tune to efficiently manage and execute such a large number of experiments, allowing for dynamic resource allocation and parallel execution across multiple processes.
\Cref{fig:pl_success_rate} compares the placement success rate across different topologies (hierarchical, random, and star) under various placement strategies (best-fit, first-fit, and min-energy) and update policies (degrade and random kill).

The results show that the \cd{first-fit} strategy achieves the highest placement success rate. The \cd{min-energy} strategy ranks second but reduces success to nearly half of the \cd{first-fit}'s performance. The \cd{best-fit} strategy performs poorly since selecting nodes approaching full capacity limits future placements and increases the risk of failure. Similarly, centralised bottlenecks at the hub node lead to worse performances in the star topology. 
With the \cd{degrade(X)} policy, a higher \cd{X} slightly improves the success rates, never exceeding 15\% and highlighting the limitations of cumulative resource reductions. In contrast, the \cd{kill(X)} policy shows declining success rates as \cd{X} increases, but maintains better resilience by continuing placements despite failures.

Overall, the optimal setup for maximum placement success combines the \cd{first-fit} strategy with a medium-to-high \cd{random} topology and \cd{kill(X)} policy with low probability.

%% file: tab/grid-params.tex
\begin{table}[t]
\renewcommand{\arraystretch}{1.2}
\centering
\caption{Grid evaluation parameters.}
\label{tab:grid_params}
\begin{tabular}{|l|l|}
\hline
\textbf{Parameter}              & \textbf{Values}                  \\ \hline
\textbf{Initial load}           & 0\%, 25\%, 50\%, 75\%            \\ \hline
\textbf{Network topologies}     & Star, Hierarchical, Random       \\ \hline
\textbf{Network sizes}          & 50, 100, 300 nodes               \\ \hline
\textbf{Placement strategies}   & first-fit, best-fit, min-energy  \\ \hline
\textbf{Update policies}        & degrade(X), kill(X)              \\ \hline
\textbf{Simulation duration}    & 600 ticks                        \\ \hline
\end{tabular}
\end{table}

%% file: src/experiment/user-distribution.tex
\begin{figure*}[t]
    \centering
    \includegraphics[width=.95\textwidth]{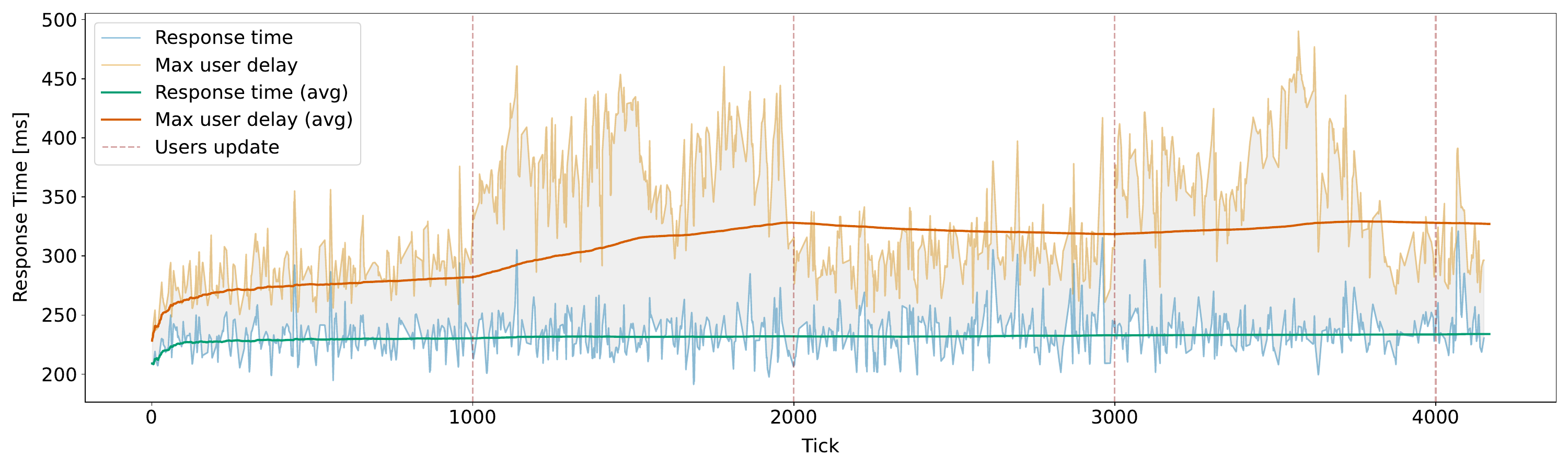}
    \caption{Response time and maximum user delay of the SockShop application.}
    \label{fig:response_time_variation}
\end{figure*}

\begin{figure}[t]
    \centering
    \includegraphics[width=0.99\linewidth]{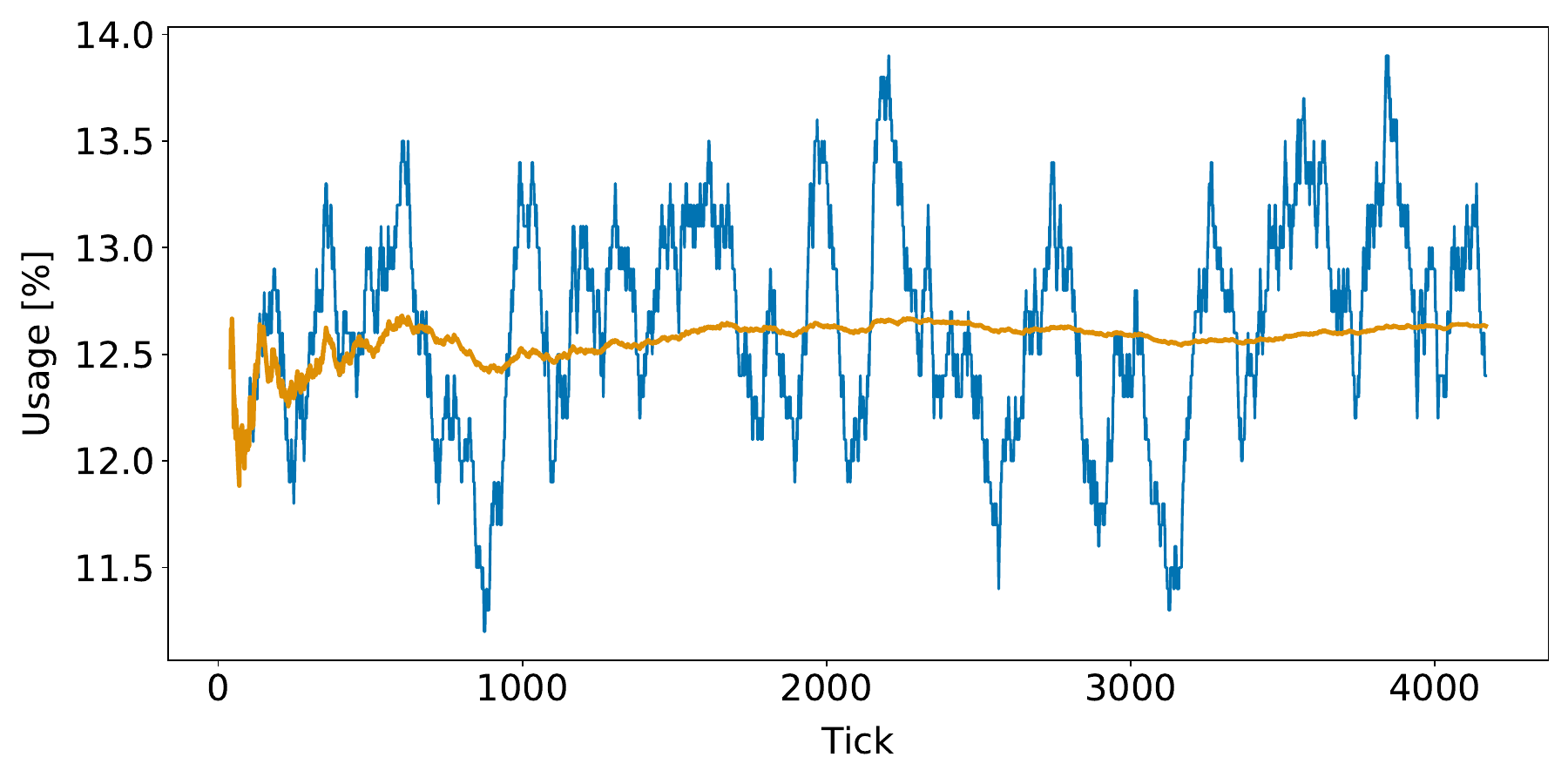}
    \caption{CPU Usage for UC2.}
    \label{fig:cpu_usage_2}
\end{figure}

\subsection{Response time analysis with varying user distribution}
\label{sec:usecase2}

In this use case, an e-commerce company wants to monitor user experience during global sales campaigns by dynamically adapting workload distribution based on fluctuating user traffic over time. Thus, such a company developing and operating a distributed application aims to analyse how its application's response time varies as users' distribution changes across the network. Using the \eclypse framework, the company can model scenarios where user traffic surges in specific areas or periods, enabling it to predict how these variations affect response times and network performance.

To do so, we employed a real user distribution based on the dataset \cd{DATA7}~\cite{data7}, which was developed to simulate mobility patterns and edge workloads in an urban area. \cd{DATA7} merges synthetic vehicle trajectories generated by the SUMO traffic simulator with real cellular tower positions (representing edge devices) from OpenCelliD\footnote{\url{https://opencellid.org/}}. \cd{DATA7} contains data on 187 edge devices and the distribution of about 3000 users (vehicles) across those devices.

\subsubsection{Simulation Environment}
To model this scenario, the simulation environment consisted of the distributed e-commerce application, the infrastructure, and the dynamic distribution of users. 
We selected the SockShop microservice-based application\footnote{\url{https://github.com/ocp-power-demos/sock-shop-demo}} -- a standard benchmark for e-commerce systems -- to model the distributed application to monitor.
We used a hierarchical infrastructure of 187 nodes, reflecting the edge devices reported in the \cd{DATA7} dataset.

\subsubsection{Simulation Workflow}
The event set $\Evt$ includes two update policies: \cd{kill(X)} for random node failures and recovery, and \cd{user-update}, which dynamically adjusts user counts on nodes based on the dataset \cd{DATA7}. 
Every 1000 ticks, the system alternates between user-doubling and user-halving events to simulate realistic workload fluctuations, such as peak and off-peak periods. The \cd{first-fit} strategy places services on the first node that meets resource requirements.

Callbacks $\Cbk$ track the \textit{user count} on each node and compute the \textit{user delay}. The system determines the delay for each node $N$ using the following formula:
\begin{equation*}
delay(N) = lat(N, H) + uc(N) \times \log(1 + uc(N))
\end{equation*}

\noindent where $lat(N, H)$ is the network latency between node $N$ and node $H$, and $uc(N)$ is the number of users on node $N$.

\subsubsection{Results}
\Cref{fig:response_time_variation} illustrates the variation in response time for the SockShop application in relation to dynamic user distribution events over 4167 ticks. We recall that at ticks 1000 and 3000, the number of users at each node is doubled, whereas at ticks 2000 and 4000, the user count is halved.
The graph shows the application's response time alongside the average additional user delay. Initially, the overall response time stabilises, and noticeable spikes occur after the user-doubling events, as increased traffic strains the system and leads to higher delays. In contrast, halving the number of users reduces the response time, although not as significantly as the spikes from user surges. This means that while the system maintains stable performance under typical conditions, it becomes more strained during sudden workload increases, reflecting the need for further optimisation to efficiently handle such dynamic changes in user distribution.

\subsubsection{Performance Evaluation}
According to the recorded metric, the simulation lasted on 1630 seconds ($\simeq$ 27 min) on average, achieving a rate of 2.5 ticks/s. Memory usage remained stable at 775 MB, reflecting the simulator's need to instantiate all required components, including the placement strategy, update policy (and associated dataset), and the metrics to be monitored, each contributing to the baseline memory consumption.
\eclypse CPU usage fluctuated between 10 and 14\%, as shown in \cref{fig:cpu_usage_2}. Although this percentage is relatively low compared to the total CPU capacity, the nominal consumption is high due to the absence of a waiting period between ticks, which accelerates the overall simulation time.

%% file: src/experiment/image-prediction.tex
\begin{figure*}[t]
    \centering
    \includegraphics[width=.95\linewidth]{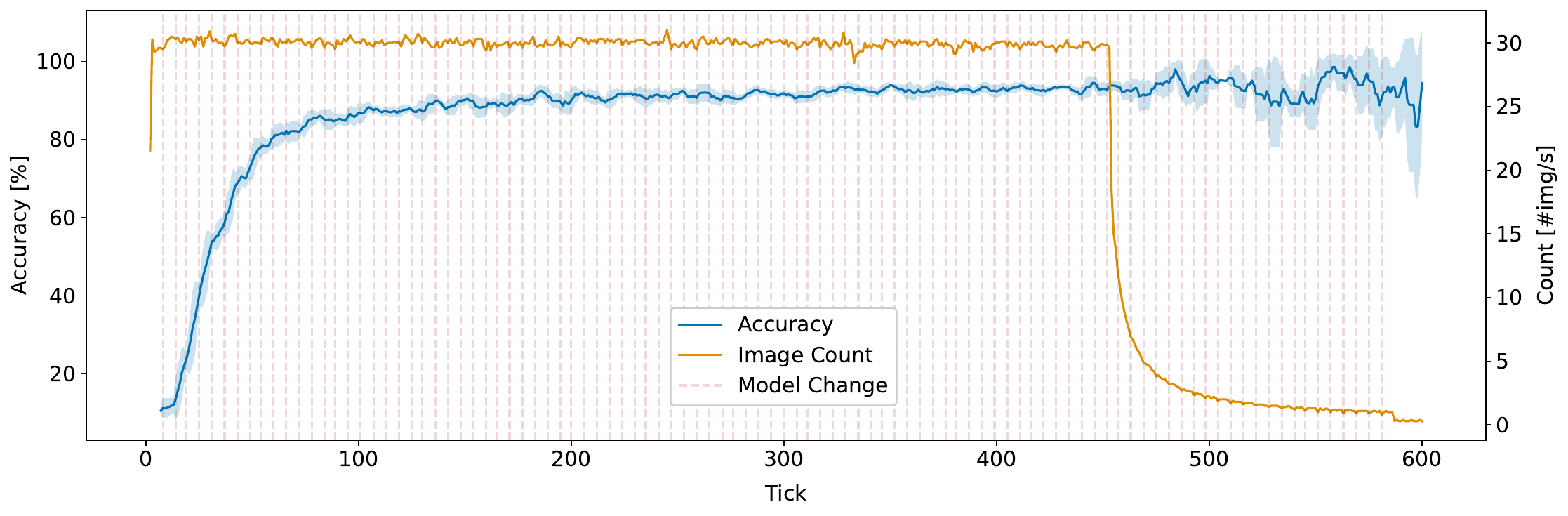}
    \caption{Trade-off between number of images exchanged per tick \textit{(1)}, and average prediction accuracy \textit{(2)}. The shaded area denotes the standard deviation.}
    \label{fig:accuracy}
\end{figure*}

\begin{figure}[t]
    \centering
    \includegraphics[width=0.95\linewidth]{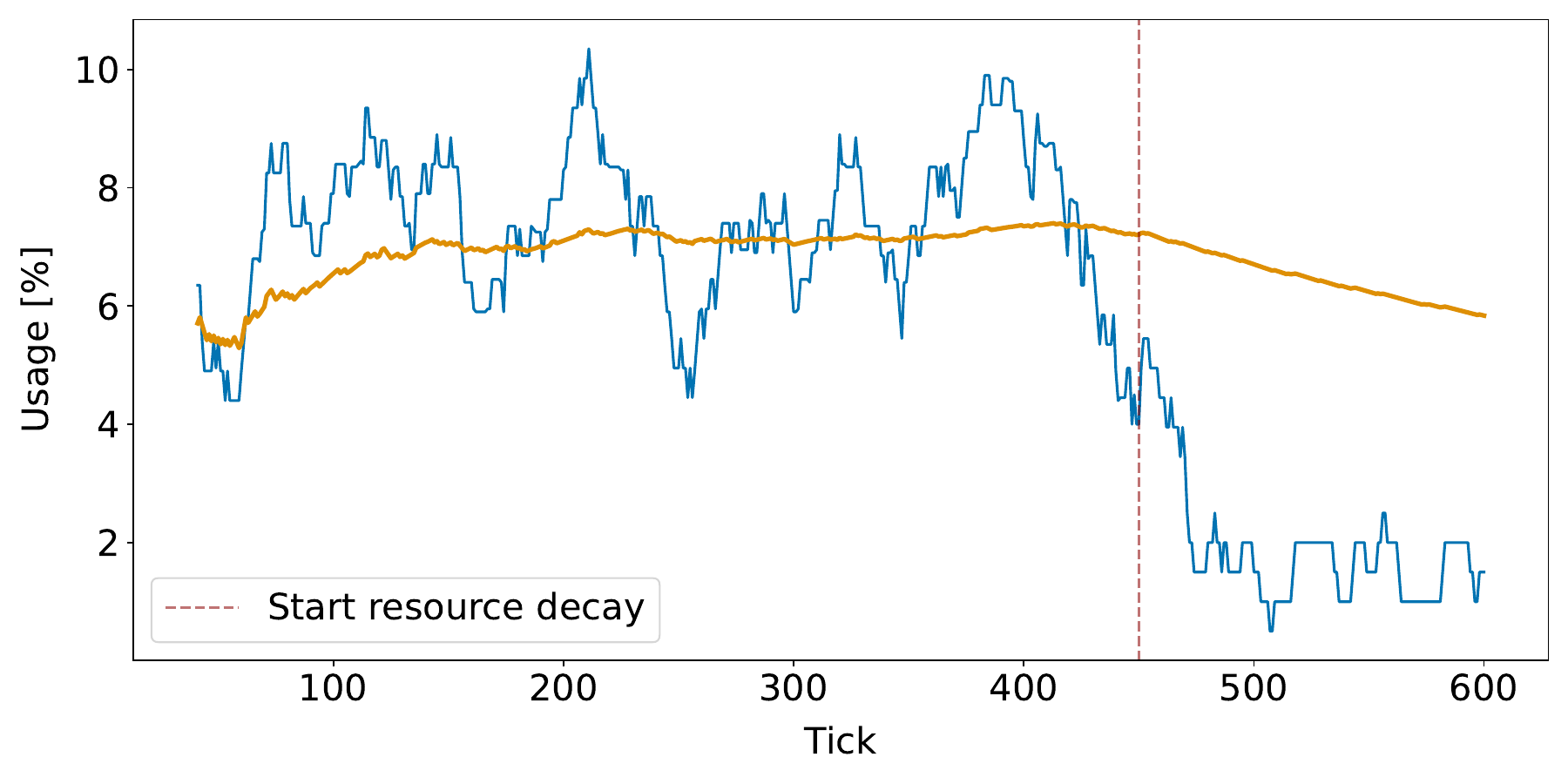}
    \caption{CPU usage during UC3 emulation, showing fluctuations as services exchange messages. A drop after tick 450 reflects reduced message exchanges due to increased network latency from a link failure policy.}
    \label{fig:cpu_usage_3}
\end{figure}

\subsection{Computer vision application}
\label{sec:usecase3}

This last use case implements a scenario that emulates an application leveraging machine learning functionalities. In particular, the given application consists of three services:
\begin{enumerate*}[label=(\arabic{*})]
    \item the \cd{EndService} implements the frontend for the end-users and streams images that require a prediction to the \cd{ImagePredictor} service;
    \item the \cd{ImagePredictor} service integrates a neural network in inference mode and, by exposing a REST endpoint \cd{/predict}, performs and returns predictions on the images it receives;
    \item the \cd{Trainer} service encapsulates a neural model with the same architecture of the \cd{ImagePredictor}, but in training mode, and sends a new version to the \cd{ImagePredictor} after every training epoch.
\end{enumerate*}
Periodically, \cd{ImagePredictor} changes the underlying model and starts predicting with its new version provided by \cd{Trainer}.

Unlike previous use cases, the simulation proceeds through a periodic trigger $\trg \in \Trg$, which occurs every 10 seconds, to provide enough time for the application to run.
All machine learning functionalities were implemented in PyTorch~\cite{pytorch}.

The set of events $\Evt$ also includes: a static placement strategy, as assessing the placement quality is not in the scope of this use case; an update policy implementing a link failure at tick 450, resulting in the degradation in the infrastructure's latency. 

Finally, we integrated the callbacks $\cbk \in \Cbk$ to account for two remote metrics (\ie that are operated by remote services):
\begin{enumerate*}[label=\textit{(\arabic{*})}]
    \item \cd{image\_count} determines the number of images processed by the predictor during each tick;
    \item the predictor's average \cd{accuracy} in the same interval.
\end{enumerate*}

\subsubsection{Results}
In \cref{fig:accuracy},  we show the metric assessing the quality of the functional aspects of the application. We can observe that accuracy grows over time during the first phase, proving the effectiveness of the training functionalities and a solid replication of the A/B testing functionalities in real-world applications. Then, after the link degradation at tick 450, we can observe an interesting behaviour that further validates the simulation quality. As the number of images exchanged decreases as a result of the higher latency, we can see that the accuracy becomes less stable. This is because the number of images on which the metric is computed is lower and the sample of images sent to the predictor exposes a more substantial bias because of the lower sample size.

In general, we can conclude that the emulation functionalities can closely replicate the behaviour of applications deployed in real-world infrastructure, and allow to foresee dynamics that would not arise through a simple static analysis.

\subsubsection{Performance Evaluation}
As expected, the emulation lasted 6000 seconds (100 min), with a rate of 0.1 ticks/s. Memory usage increased by only 10 MB (from 650 to 660 MB) during the simulation, slightly lower than in the previous case due to the reduced number and lighter weight of components managed by the simulator. CPU usage, as illustrated~\cref{fig:cpu_usage_3}, directly reflects the dynamic behaviour of the emulation. 
Messages exchanged among emulated services are subject to delays determined by link latencies, computed and updated by the simulator. Thus, when nodes need to exchange messages, they ask for routing paths to the simulator. However, as latencies increase after 450 ticks, messages are sent less frequently due to the degradation policy. Consequently, requests for routes -- and the corresponding computations and transmissions by the simulator -- occur less often, leading to less resource usage.

%% file: src/related.tex
\section{Related Work}
\label{sec:related}

\noindent
Since Cloud-Edge paradigms have been introduced, many studies have addressed the problem of simulating or emulating Cloud-Edge environments, modelling various application and infrastructure characteristics~\cite{survey2019FI,survey2020_DeMaio,survey2020_Kertesz} and featuring different emulation or simulation capabilities to assess application and resource management policies. 
This section reviews existing Cloud-Edge system simulation and emulation work, highlighting how \eclypse advances the state of the art. Results of our analyses are summarised in~\cref{tab:simulators}.

Among the first proposals for simulating Cloud-Edge scenarios, \textit{iFogSim}~\cite{iFogSim2017} is a Java framework based on \textit{CloudSim}~\cite{CloudSim}. iFogSim models hierarchical (tree-shaped) Cloud-Edge infrastructure topologies with IoT sensors and actuators and processing and networking capabilities. Applications are modelled as direct acyclic workflows of tasks, following a sense-process-actuate model. iFogSim can configure CPU, RAM, MIPS, uplink and downlink bandwidth, and busy/idle power to estimate energy consumption. iFogSim does not support mobility or VM/container migration. More recently, \textit{iFogSim2}~\cite{iFogSim2} extends the previous version of the simulator with mobility aspects, resource clustering features, and microservice-based applications.  
Building on top of iFogSim, \textit{MobFogSim}~\cite{MobFogSim} specifically allows simulating user movements and assessing their distance from deployed services to decide their placement dynamically.

Still based on CloudSim and written in Java, \textit{EdgeCloudSim}~\cite{EdgeCloudSim} and, more recently, \textit{PureEdgeSim}~\cite{PureEdgeSim} are prototype simulators designed to evaluate the performance of cloud-edge computing systems, considering both computational and network resources. EdgeCloudSim can specifically model device mobility, WLAN and WAN networks, and edge servers without considering application workload migration. 
PureEdgeSim, on the other hand, is designed to handle scenarios with many heterogeneous devices and different resource management strategies. It supports sense-process-actuate and stream-processing application models and accounts for mobility and battery energy consumption. 
Similarly, \textit{IoTSim-Edge}~\cite{IoTSim-Edge} and \textit{IoTSim-Osmosis}~\cite{IoTSim-Osmosis} extend CloudSim, focussing on data offloading from IoT devices to edge datacentres and on osmotic computing applications, respectively. While IoTSim-Edge focuses on device mobility and device-to-device interactions, IoTSim-Osmosis mainly considers Cloud-Edge workload migration over static infrastructure conditions.

Other solutions focusing on QoS awareness include \textit{FogTorch$\Pi$}~\cite{FogTorchPI} and \textit{RECAP}~\cite{RECAP2017}, tools for Cloud-Edge application deployment decisions based on QoS, costs, and resource use in dynamic settings. The former relies on a stochastic infrastructure model that accounts for resource availability variations (\eg failures, workload changes, mobility) and includes RAM, CPU, hard disk, IoT devices, and network QoS. Using backtracking or genetic algorithms, Monte Carlo simulations assess eligible application placements against infrastructure variations, with a multithreaded Java prototype supporting their execution. Similarly, RECAP was intended to enable reproducible and configurable resource and application management assessment in Cloud-Edge landscapes accounting for QoS, energy efficiency, and costs.

Being among the first Cloud-Edge simulation environments written in Python, \textit{YAFS}~\cite{YAFS} retakes and extends the model of iFogSim~\cite{iFogSim2017} by allowing simulating arbitrary infrastructure topologies, dynamic application and infrastructure management policies, node mobility and infrastructure dynamicity (\eg node/link failures). Unlike previous efforts, it enables using simulation traces to ease experimental reproducibility and data analysis. YAFS has also recently been used to simulate decentralised application and infrastructure management~\cite{MARIO3}. 
Still written in Python, \textit{FogDirSim}~\cite{FogDirSim} relies on formal modelling of Cisco FogDirector's RESTful API to simulate the lifecycle management of Cloud-Edge applications by comparing and contrasting alternative management policies against stochastic network and workload variations. 
Recently, with a more general aim, \textit{EdgeSimPy}~\cite{edgesimpy} also allows its users to model and assess resource management policies in Cloud-Edge environments. It models the lifecycle and management of container-based applications, including provisioning, migration, updates, user mobility, and energy consumption.

Speaking of emulation, to the best of our knowledge, only three distinct frameworks have been proposed so far for emulating Cloud-Edge systems. 
\textit{EmuFog}~\cite{emufog} permits emulating large infrastructures, emphasising scalability and extensibility. It relies on Docker and MaxiNet~\cite{maxinet}, which enables distributed emulations and is designed to test resource management policies and applications in realistic conditions before deployment, considering latency and cost characteristics. Besides emulation, the EmuFog workflow includes topology generation, transformation, and extension.
On a similar line, \textit{FogBed}~\cite{fogbed} is an emulation framework oriented towards rapidly prototyping Cloud-Edge components. It relies on Mininet~\cite{mininet} and Docker to emulate the nodes, creating a low-cost, flexible, and real-world test environment. Resource models allow emulating the CPU, memory, and storage limitations typical of edge devices. The Fogbed workflow includes defining container images, configuring the topology, starting the emulator, and managing instances. Last, \textit{Fogify}~\cite{fogify} allows configuring and emulating Cloud-Edge topologies, including heterogeneous resource characteristics, network capabilities, and QoS criteria and allows for injecting errors and changing the configuration at runtime. Fogify offers a Python SDK for interacting with the emulated environment and collecting performance, QoS and cost metrics. 
Recently, \textit{iContinuum}~\cite{icontinuum} was proposed as an emulation toolkit that specifically targets intent-based computing in software-defined networking settings over the Cloud-Edge continuum. By leveraging Kubernetes, ONOS software-defined capabilities and Mininet for network emulation, it allows replicating deployment scenarios to enforce high-level intent objectives, \eg optimising application response time or energy consumption. 

\input{tab/related-comparison}

\Cref{tab:simulators} summarised the features of each framework over ten key dimensions: the \textit{programming language} used, support for \textit{dynamic environments} (\ie time-varying conditions), \textit{mobility} of users/services, support for \textit{event-driven workflows}, availability of structured \textit{reporting}, \textit{emulation} capabilities, \textit{extensibility} of core components, and quality of \textit{documentation}.

Legacy tools such as \textit{CloudSim} and its early extensions (\eg \textit{iFogSim2}, \textit{EdgeCloudSim}) primarily focus on hierarchical infrastructures with static behaviours and limited extensibility. 
More recent efforts like \textit{CloudSim 7G}~\cite{andreoli2025cloudsim} improve modularity and integration capabilities, but remain restricted to Java-based environments with no support for emulation or runtime dynamics.
Python-based simulators such as \textit{YAFS}, \textit{EdgeSimPy}, and \textit{FogDirSim} offer greater flexibility and are closer in spirit to \eclypse, but still lack comprehensive support for graph-based execution logic, asynchronous event workflows, and native integration of emulated components.

Among emulation platforms, tools like \textit{EmuFog} and \textit{Fogify} focus on Docker- and Mininet-based infrastructure realism but miss the abstraction and generality of simulation. iContinuum~\cite{icontinuum} enables intent-based orchestration, but remains tightly bound to SDN technologies.

In contrast, \eclypse uniquely blends simulation and emulation under a unified event-driven model, offering maximum flexibility, full Python interoperability, and a fully customisable graph-based execution environment with rich documentation and extensibility features. Unlike many of the surveyed tools -- which often suffer from incomplete documentation, limited maintainability, or runtime instability~\cite{usability} -- \eclypse is distributed as an official Python module through the \cd{pip}\footnote{\url{https://pypi.org/project/eclypse/}} package manager and is supported by extensive documentation and ready-to-use examples.

Moreover, being fully embedded within the Python ecosystem, \eclypse enables seamless integration with state-of-the-art tools and libraries, such as PyTorch~\cite{pytorch}, TensorFlow~\cite{tensorflow} and scikit-learn~\cite{scikit-learn} for machine learning-based decision models, NetworkX and Matplotlib\footnote{\url{https://matplotlib.org}} for custom topological analysis and visualisation, and Jupyter\footnote{\url{https://jupyter.org}} for interactive, reproducible experiments. This compatibility facilitates the development of advanced Cloud-Edge workflows and also accelerates experimentation cycles through rapid prototyping and analysis.

%% file: tab/related-comparison.tex
\begin{table*}[t]
	\caption{Comparison of \eclypse with state-of-the-art Cloud-Edge simulation and emulation frameworks.\label{tab:simulators}}
	\centering
	\begin{tabular*}{\textwidth}{@{\extracolsep\fill}l|cccccccccc@{}}
		\toprule
		\textbf{Framework} & \textbf{Language} & \makecell{\textbf{Dynamic}\\\textbf{Environment}} & \textbf{Mobility} & \textbf{Events} & \textbf{Reporting} & \textbf{Emulation} & \textbf{Extensibility} & \textbf{Docs} \\
		\midrule
		\eclypse & Python & \checkmark & \checkmark & \checkmark & \checkmark & \checkmark & \checkmark & \checkmark \\ \midrule
		\cite{iFogSim2017} iFogSim & Java & \texttimes & \texttimes & $\triangle$ & \texttimes & \texttimes & $\triangle$ & \texttimes \\ \hline
		\cite{iFogSim2} iFogSim2 & Java & \checkmark & $\triangle$ & $\triangle$ & \texttimes & \texttimes & \checkmark & \texttimes \\ \hline
		\cite{MobFogSim} MobFogSim & Java & \texttimes & \checkmark & $\triangle$ & \texttimes & \texttimes & \texttimes & \texttimes \\ \hline
		\cite{EdgeCloudSim} EdgeCloudSim & Java & $\triangle$ & \checkmark & $\triangle$ & \texttimes & \texttimes & $\triangle$ & \texttimes \\ \hline
		\cite{PureEdgeSim} PureEdgeSim & Java & \checkmark & \checkmark & \checkmark & \texttimes & \texttimes & \checkmark & \texttimes \\ \hline
		\cite{IoTSim-Edge} IoTSim-Edge & Java & \checkmark & \checkmark & \checkmark & \texttimes & \texttimes & $\triangle$ & \texttimes \\ \hline
		\cite{IoTSim-Osmosis} IoTSim-Osmosis & Java & \checkmark & \texttimes & \checkmark & \texttimes & \texttimes & $\triangle$ & \texttimes \\ \hline
		\cite{FogTorchPI} FogTorch$\Pi$ & Java & $\triangle$ & \checkmark & $\triangle$ & \texttimes & \texttimes & $\triangle$ & \texttimes \\ \hline
		\cite{RECAP2017} RECAP & Java & \checkmark & \texttimes & \checkmark & \texttimes & \texttimes & \checkmark & \texttimes \\ \hline
        \cite{andreoli2025cloudsim} CloudSim 7G & Java & $\triangle$ & \texttimes & \checkmark & \checkmark & \texttimes & \checkmark & $\triangle$ \\ \hline
		\cite{YAFS} YAFS & Python & \checkmark & \checkmark & \checkmark & \texttimes & \texttimes & \checkmark & $\triangle$ \\ \hline
		\cite{FogDirSim} FogDirSim & Python & \checkmark & \checkmark & \checkmark & $\triangle$ & \texttimes & \checkmark & \texttimes \\ \hline
		\cite{edgesimpy} EdgeSimPy & Python & \checkmark & \checkmark & \checkmark & $\triangle$ & \texttimes & \checkmark & \checkmark \\ \hline
		\cite{fogify} Fogify & Python + Docker & \checkmark & \checkmark & \checkmark & \checkmark & \checkmark & \checkmark & $\triangle$ \\ \hline
		\cite{emufog} EmuFog & Python + Docker & \texttimes & \texttimes & \texttimes & \texttimes & \checkmark & \checkmark & \texttimes \\ \hline
		\cite{fogbed} FogBed & Python + Mininet & \texttimes & \texttimes & \texttimes & \texttimes & \checkmark & \checkmark & \texttimes \\ \hline
		\cite{icontinuum} iContinuum & Python + K8s & $\triangle$ & \checkmark & \checkmark & \checkmark & \checkmark & \checkmark & $\triangle$ \\
		\bottomrule
	\end{tabular*}
	\begin{tablenotes}
		\item[] \textbf{Legend:} \checkmark = supported, \texttimes\, = not supported, $\triangle$ = partial or conditional support.
	\end{tablenotes}
\end{table*}

%% file: src/conclusions.tex
\section{Concluding Remarks}
\label{sec:conclusions}

\noindent
In this article we introduced \eclypse, a Python-based framework to simulate and emulate Cloud-Edge computing environments.
\eclypse provides a flexible and versatile environment for testing application deployment strategies, optimising resource allocation, and validating application performance under dynamic, realistic conditions.
Its architecture is designed with extensibility at its core: each simulation and emulation component can be independently configured or extended, enabling users to tailor infrastructure, applications, and behaviours to specific research needs.
This high degree of customisability is made possible by the modular structure of the framework and its event-driven execution model, which together support the integration of user-defined deployment strategies, workload scheduling policies, and service-level behaviours.
As a result, \eclypse can effectively adapt to evolving Cloud-Edge paradigms and serve as a robust foundation for experimentation in both simulated and emulated environments.

In our future work, we aim at improving the simulation functionalities by enhancing network transmission logging to analyse more complex network behaviours and their impact on application performance. Additionally, we will provide a standardised benchmark factory for easier comparison with other approaches. Finally, we will provide a GUI and integrate the interaction of services with the simulation engine, making it a viable assessment tool for production environments. Furthermore, we plan to integrate \eclypse with real-world testbeds and federated infrastructures, enabling the deployment of Cloud-Edge applications in live heterogeneous environments. This will allow validation of orchestration strategies under realistic hardware and network conditions, bridging the gap between simulation, emulation, and real execution.

%% file: src/acknowledgments.tex
\bmsection*{Acknowledgments}
{
\linespread{1}\selectfont
\noindent Work partly funded by projects: \textit{NOUS} HORIZON-CL4-2023-DATA-01-02 project, GA n. 101135927; 
\textit{TEACHING} project funded by the EU Horizon 2020, GA n. 871385;
\textit{Energy-aware management of software applications in Cloud-IoT ecosystems} (RIC2021\_PON\_A18) funded over ESF REACT-EU resources by the Italian Ministry of University and Research through \textit{PON Ricerca e Innovazione 2014--20};
\textit{OSMWARE} project (PRA 2022-64), funded by the University of Pisa;
“SEcurity and RIghts In the CyberSpace - SERICS” (PE00000014 - CUP H73C2200089001) under the National Recovery and Resilience Plan (NRRP) funded by the European Union - NextGenerationEU.
\par}